\documentstyle[12pt]{article}

\begin {document}
~\hspace*{5.0cm} ADP-AT-96-1\\
~\hspace*{5.9cm} {\it Astroparticle Physics}, in press\\

\begin{center}
{\large \bf A new estimate of the extragalactic radio background and
implications for ultra-high-energy gamma-ray propagation}\\[1cm]
R.J. Protheroe$^1$ and P.L. Biermann$^2$\\ $^1$Department of Physics and
Mathematical Physics\\ The University of Adelaide, Adelaide, Australia
5005\\ $^2$Max-Planck Institut f\"{u}r Radioastronomie, Auf dem H\"{u}gel
69, D-53121 Bonn, Germany\\[2cm]
\end{center}

\begin{center}
\underline{Abstract}\\
\end{center}

We make a new estimate of the extragalactic radio background down to kHz
frequencies based on the observed luminosity functions and radio spectra of
normal galaxies and radio galaxies.  We have constructed models for the
spectra of these two classes of objects down to low frequencies based on
observations of our Galaxy, other normal galaxies and radio galaxies.
We check that the models and evolution of the luminosity functions give
source counts consistent with data and calculate the radio background
expected from kHz to GHz frequencies.

The motivation for this calculation is that the propagation of ultra-high
energy gamma-rays in the universe is limited by photon-photon pair
production on the radio background.  Electromagnetic cascades involving
photon-photon pair production and subsequent synchrotron radiation in the
intergalactic magnetic field may develop.  Such gamma-rays may be produced
in acceleration sites of ultra-high energy cosmic rays, as a result of
interactions with the microwave background, or emitted as a result of
decay or annihilation of topological defects.  We find that photon-photon
pair production on the radio background remains the dominant attenuation
process for gamma-rays from $3 \times 10^{10}$ GeV up to GUT scale
energies.

\section{Introduction}

The universe is not transparent to high energy gamma-rays due to 
interactions with low energy photons of the extragalactic radiation
fields, the most important process being photon-photon pair production.
For example, interactions in the cosmic microwave background radiation
give a mean interaction length of less than 10 kpc at $10^6$ GeV as
has been known since soon after the discovery of the microwave background 
\cite{Gould,Jelley}.
The threshold for interactions on the microwave background is 
$\sim 10^5$ GeV, and at lower energies interactions on the infrared
and optical backgrounds limit the transparency at TeV energies 
(e.g. \cite{StdeJS,Pro93})
Other components of the extragalactic
background radiation are discussed in the review of Ressel and Turner
\cite{Res90}.

Above $\sim 10^{10}$ GeV interactions with the radio background
become more important than the microwave background in limiting the
transparency of the universe to gamma-rays and controlling any 
resulting electromagnetic cascades.
Both the infrared and radio backgrounds are poorly known due to
our location within our Galaxy which emits and absorbs at these wavelengths.
The radio background was measured over twenty-five years ago \cite{Cla70}, but
the fraction of this radio background which is truly extragalactic, and not
contamination from our own Galaxy, is still debatable.  A theoretical
estimate \cite{Ber69} was made about the same time which gave a quite
different spectrum, particularly at low frequencies.  In recent
cascade calculations \cite{ProtJohns95,Elb95,Lee96,ProtStan96} the estimate of
ref.~\cite{Cla70} has been used.  It is this very uncertain radio
background which will provide target photons for UHE $\gamma$-rays above
$\sim 10^{10}$ GeV.

While gamma-ray astronomy is not currently undertaken at $\sim 10^{10}$ GeV 
energies, it is important to know the photon-photon mean interaction length
at these energies because cascading involving gamma-rays at these 
energies occurs in top-down models for the origin of the highest energy
cosmic rays.
The highest energy cosmic rays have energies of 200 EeV \cite{Hay94,Yos95}
and 300 EeV \cite{Bir95}, and are well above the 
``Greisen-Zatsepin-Kuzmin cut-off'' \cite{Gre66,Zat66}
at 50 EeV in the spectrum of cosmic ray protons 
due to pion photoproduction in the microwave background,
which is expected if the cosmic rays  
originate further than a few tens of Mpc ({\it e.g.}, \cite{GrigBer,RB93}).
Of the various models proposed to account for the origin of these high energy
cosmic rays \cite{Biermann&Strittmatter,RB93,Mil95,Wax95,Vie95},
one of the more tantalizing speculations is that the
highest energy cosmic rays may be ultimately due to the decay of 
supermassive X particles
\cite{Bha92,Sig94,Bha95,Sig95}, themselves radiated
during collapse or annihilation of topological defects, remnants of an
early stage in the evolution of the Universe.  The X particles have
GUT-scale masses of $\sim 10^{16}$ GeV or lower, depending on 
the theory, and decay into
leptons and quarks at lower energies.  The quarks themselves fragment into
a jet of hadrons which, it is supposed, could produce the highest energy
cosmic rays, although there is some debate as to whether a sufficiently
large fraction of the energy of the defect could end up in high energy
particles \cite{Hin95}.  In any case, much of the radiation is likely to
emerge in the electromagnetic channel and initiate an electromagnetic
cascade in the ambient radiation field, in which collisions with radio
photons play an important role.

In this paper we make a new calculation of the extragalactic radio
background down to kHz frequencies based on the infrared luminosity
function of normal galaxies recently determined from IRAS source counts,
the observed radio--infrared correlation, and the luminosity function of
radio galaxies, together with recent models for radio spectra of these
objects.  Finally, we calculate the mean free path for $\gamma$-rays in
the extragalactic radio background radiation.

\section{Calculation of radio background}

The main contributions to the radio background will be from normal
galaxies and radio galaxies and we will discuss the radio spectra of
these objects down to kHz frequencies and construct
models for their spectra.
Using appropriate luminosity functions, we will then integrate over luminosity
and redshift to obtain the radio background.
This is very sensitive to the evolution of galaxies and we shall use
radio and infrared source counts to constrain the evolution.
We start by discussing the how the
observed radio flux is related to the luminosity in an expanding
universe.

The observed flux $S_\nu$ at frequency $\nu$ is related to the luminosity at
frequency $\nu^\prime = (1+z)\nu$ by

\begin{equation}
S_\nu = {L_{\nu^\prime} \over 4 \pi d_L^2} {d \nu^\prime \over d \nu} =
 {L_{\nu^\prime} (1+z) \over 4 \pi d_L^2}
\end{equation}
where $S_\nu$ has units of W Hz$^{-1}$ m$^{-2}$,
$L_\nu$ has units of W Hz$^{-1}$,
$d_L=(1+z)d_0$ is the luminosity distance,
and $d_0$ is the physical distance at photon reception (m).

\subsection{Normal galaxies}

In calculating the radio flux of
normal galaxies, we shall use the 60 micron luminosity, $L_{60}$,
and an observed correlation between the luminosities at 1.4 GHz, 
$L_{1.4}$, and at 60 micron.
We may then write

\begin{equation}
S_\nu = {(L_{\nu^\prime}/L_{1.4})(L_{1.4}/L_{60})L_{60} (1+z)
\over 4 \pi d_L^2}.
\end{equation}

We use the observed correlation between the luminosities at 1.4 GHz
and at 60 micron given by Condon \cite{Condon92}

\begin{equation}
L_{1.4} = 1.69 \times 10^{-3} (2.58 + 1.67^\alpha) L_{60}
\end{equation}
where both $L_{60}$ and $L_{1.4}$ have units W Hz$^{-1}$, and
$\alpha$ is the spectral index at 60 micron.
Hacking et al. \cite{Hacking87} give the distribution of
spectral indices at 60 micron which can be approximated by a gaussian
distribution with mean, $\langle \alpha(L_{60}, z) \rangle$ given
by equation 2 of \cite{Hacking87}, and standard deviation $\sigma = 0.5$,
and we integrate over this distribution of $\alpha$ in our calculation.

To obtain the ratio $(L_{\nu^\prime}/L_{1.4})$ we need to know the
spectrum of normal galaxies in the radio region from GHz down to kHz
frequencies.
This spectrum is poorly known, and so we shall model the
spectrum based on observations
of the Galaxy and other galaxies above $\sim 50$ MHz together with
an estimate of the effect of free-free absorption.
In the Galaxy, the observed spectrum is a power-law,
$I_\nu \propto \nu^{-\alpha}$, with spectral index $\alpha\approx 0.9$
at high frequencies, and $\alpha\approx 0.4$ at low frequencies
\cite{LongairV2}.
If the emission from the Galaxy were observed from outside the Galaxy it
would be modified by free-free absorption by the warm and hot ionized
components of the interstellar medium, and by synchrotron self-absorption.
Free-free absorption by the warm
component may be expected to be patchy as based on observations of external
galaxies \cite{Biermann&Fricke77,KBS85,IsraelM90,Hummel91}.  However, this
patchy absorption is apparently not the cause of the observed downturn of
the radio spectra of galaxies, but rather the losses experienced by the
cosmic ray electrons at low energies.

The gamma-ray emission of the Galaxy demonstrates that the low energy
spectrum of cosmic ray electrons is modified by ionization and
bremsstrahlung losses below  about 400 MeV, and cuts off below
about 50 MeV  \cite{Strong96}.  Models can be constructed that  explain
both the radio  as well as the gamma-emission from the Galaxy.   Such models
then have an approximate spectrum of cosmic ray electrons as follows

\begin{equation}
n_e(E) \; \propto \; \left\{
\begin{array}{ll}
 (E/400 \; {\rm MeV})^{-2.8}  & {\rm > \; 400 \; MeV} \\
(E/400  \; {\rm MeV})^{-1.8} & {\rm < \; 400 \; MeV} \\
( \; {\rm 50 \; MeV/400 \; MeV})^{-1.8} & {\rm < \; 50 \; MeV}
\end{array} \right.
\label{eq:e_spec}
\end{equation}
There has to be a low energy
cutoff in the electron spectrum, such as exists in low energy
protons (\cite{NB94}); such a cutoff arises from the extreme losses in
ionization and heating of the interstellar medium on the one hand, but we
also need to note that energetic electrons can be accelerated in Supernova
remnant shocks only at those energies for which their Larmor radius exceeds
that of the thermal protons in a shock (\cite{Bell78}), and that corresponds
to a few tens of  MeV.  Below this energy the electrons gain and lose energy by
interaction with plasma waves (\cite{Lesch}).

\begin{figure}
\vspace{16cm}
\includegraphics{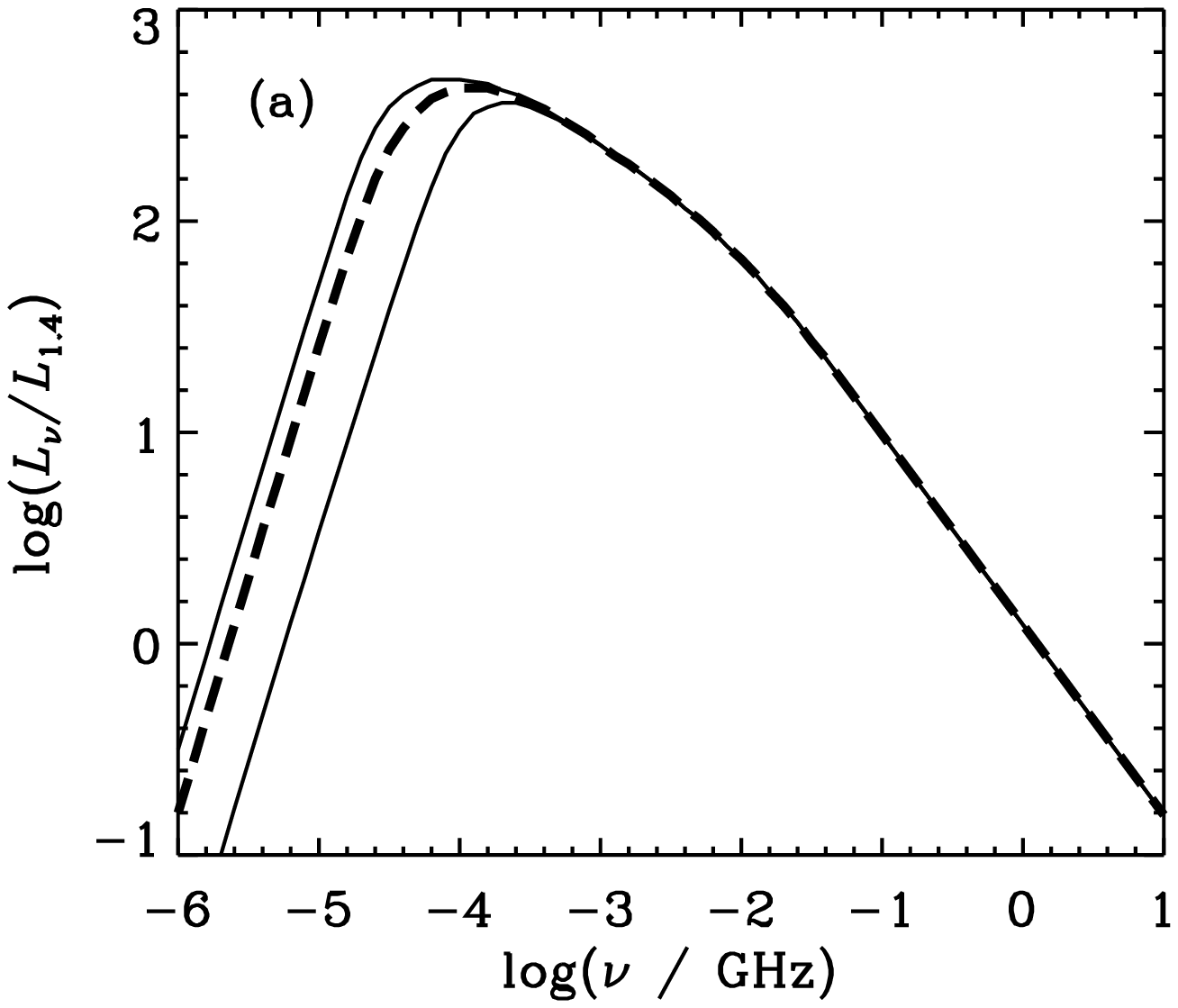}
\includegraphics{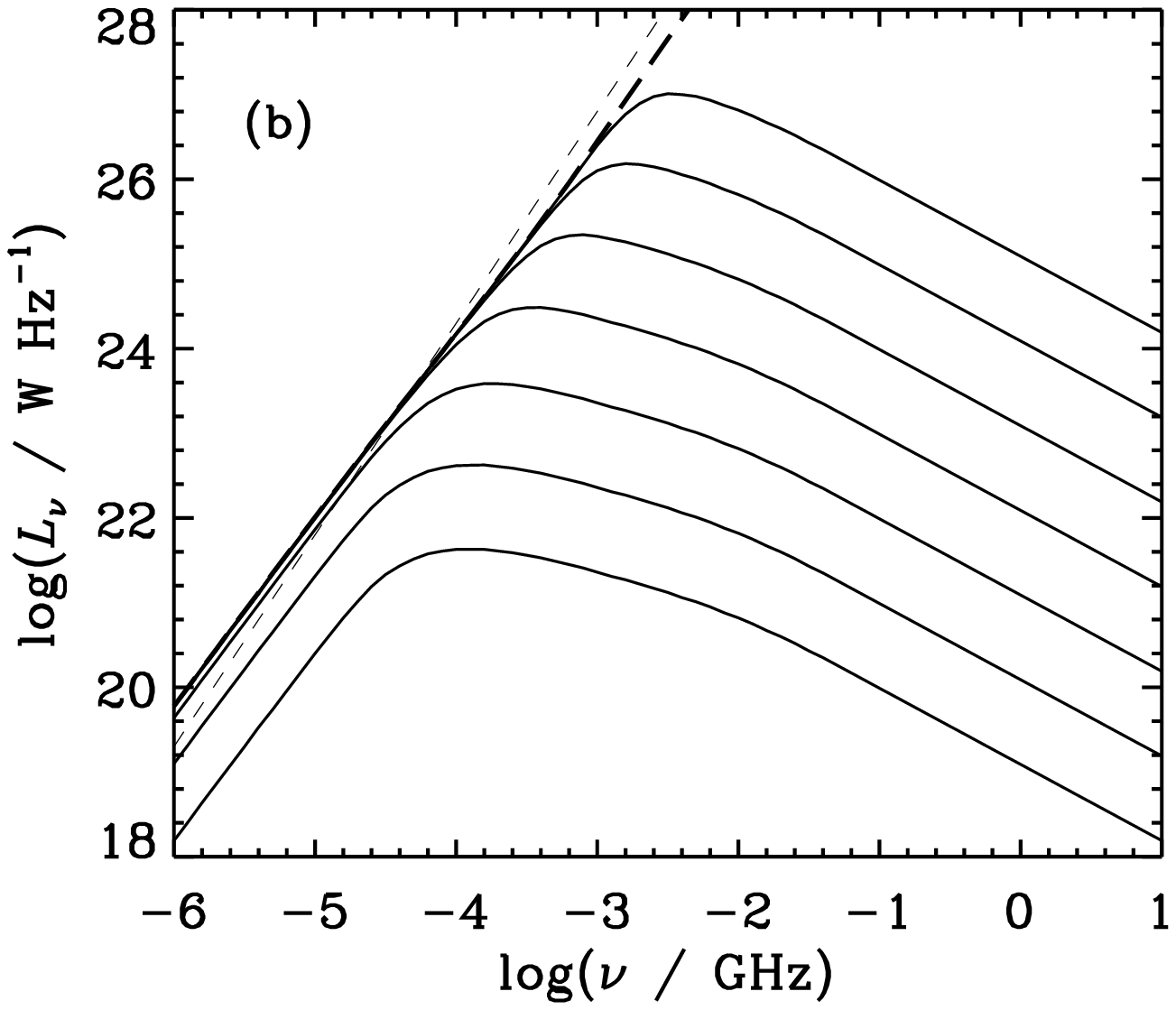}
\caption{(a) Model radio spectrum for normal galaxies neglecting
synchrotron self-absorption.
Solid lines show the range of spectra for viewing angles
$0^\circ$ (top, face-on) and $90^\circ$ (bottom, edge-on);
dashed line shows average spectrum assuming isotropic distribution of
viewing directions.
(b) Average spectra for normal galaxies with luminosity
$L_{1.4} =10^{19}$ \dots $10^{25}$ W Hz$^{-1}$ including the
effects of synchrotron self-absorption.
Dashed lines show spectrum of completely self-absorbed source:
long dashes -- accurate treatment, short dashes -- approximate treatment.}
\label{fig:freefree}
\end{figure}

We use the electron spectrum of Equation~\ref{eq:e_spec}, a magnetic field of
6 microgauss \cite{Beck96}, and apply the standard formulae for obtaining the
emission coefficient (W m$^{-3}$ Hz$^{-1}$ sr$^{-1}$) for synchrotron radiation
\begin{equation}
\varepsilon_\nu = {1 \over 2} c_3 B \langle \sin \theta \rangle
 \int_0^\infty n_e(E) F(x) dE
\end{equation}
where $x=\nu/\nu_c$,
\begin{equation}
\nu_c =
{3 \over 2} \nu_g \langle \sin \theta \rangle \gamma^2 ,
\end{equation}
$\nu_g$ is the electron cyclotron frequency,
$\langle \sin \theta \rangle = 0.785$ for isotropic electrons, and $c_3$
and the function $F(x)$ are given in ref.~\cite{Pacholczyk}.
The amount of 
absorption by the hot component of the interstellar medium will also
depend on viewing angle.
We assume a density of 0.01 cm$^{-3}$ and a scale height of 1 kpc
\cite{TaylorCordes}, a temperature of $3 \times 10^5$ K and radial extent of
15 kpc giving pathlengths through the galaxy ranging from 2 to 30 kpc.
Since the radio synchrotron radiation is observed to originate in
approximately the same volume, we solve the equation of radiative transfer
for the case of the emission and absorption coefficients
being independent of position throughout this volume.
The intensity is then
\begin{equation}
I_\nu  = {\varepsilon_\nu \over \kappa_\nu}
[1 - \exp(-\tau_\nu)],
\end{equation}
where $\varepsilon_\nu$ is the synchrotron emission coefficient
and $\kappa_\nu$ and $\tau_\nu$ are the free-free absorption coefficient
and optical depth.
For a given viewing angle we model the spectrum in this way neglecting
synchrotron self-absorption.
Assuming an isotropic distribution of viewing angles, we obtain
the average spectrum in the radio region shown
in Figure~\ref{fig:freefree}(a).

Synchrotron-self-absorption will be important at very low frequencies if
the spectrum is not cut off by other processes.
Following Longair \cite{LongairV2} the flux at frequency $\nu$ from
a self-absorbed synchrotron source may be approximated by

\begin{equation}
S_\nu \approx {2 m_e \over 3 \nu_g^{1/2}} \Omega \nu^{5/2}
\end{equation}
where $m_e$ is the electron mass
and $\Omega$ is the solid angle subtended by the source.
For a source at distance $d$ this implies a luminosity
$L_\nu = 4 \pi d^2 S_\nu$
at frequency $\nu$ and solid angle $\Omega = A_{\rm proj}/d^2$,
where $A_{\rm proj}$ is the projected area of the source normal to the line
of sight, we obtain

\begin{equation}
L_\nu \approx {2 m_e \over 3 \nu_g^{1/2}} 4 \pi A_{\rm proj} \nu^{5/2}
\label{ssa_approx}
\end{equation}
for a self-absorbed source.
For a typical galactic magnetic field
of a few microgauss (6 microgauss in the solar neighborhood \cite{Beck96})
and typical dimensions of the synchrotron emitting region of normal galaxies
(radius  $\sim 15$ kpc, height $\sim 1$ kpc) we find synchrotron
self-absorption to be important for $L_{1.4} > 10^{20}$ W Hz$^{-1}$.

Because synchrotron self-absorption will determine the shape of the
low frequency part of the spectrum of normal galaxies we now give a
more accurate treatment.
The absorption coefficient is given by
\begin{equation}
\kappa_\nu = {c^2  c_3 \over 2 \nu^2} B \sin \theta
 \int_0^\infty E^2 {d \over dE} \left[ {n_e(E) \over E^2} \right] F(x) dE,
\end{equation}
and where the source is self-absorbed the intensity will be given by
the source function
\begin{equation}
{\varepsilon_\nu \over \kappa_\nu} = { \nu^2
 \int_0^\infty n_e(E) F(x) dE \over c^2
 \int_0^\infty E^2 {d \over dE} [ {n_e(E) / E^2} ] F(x) dE} .
\end{equation}
The maximum luminosity at frequency $\nu$ is then
\begin{equation}
L_\nu = 4 \pi A_{\rm proj} {\varepsilon_\nu \over \kappa_\nu}.
\label{ssa_accurate}
\end{equation}
This is plotted  as the long-dashed line in Figure~\ref{fig:freefree}(b)
where average spectra of normal galaxies are also shown
for a range of luminosities.
The approximate result (Equation~\ref{ssa_approx}) is also shown
(short-dashed line) and is seen to give a reasonable approximation
to the more accurate treatment above.
In summary, we note that all
three effects, synchrotron self-absorption, free-free absorption in the hot
medium, and a low energy cutoff of the electron spectrum, contribute to cut
off the spectrum at kHz to MHz frequencies.

The 60 micron luminosity function, $\rho(L_{60},z)$, is the number of sources
at redshift $z$ per unit co-moving volume at 60 micron luminosity $L_{60}$
per unit of luminosity.
It has units of Mpc$^{-3}$ (W Hz$^{-1}$)$^{-1}$.
We use the local luminosity function, $\rho_0(L_{60})$, based on the
local visibility function given by Hacking et al. \cite{Hacking87}.
The local luminosity function is obtained from equation 3
of \cite{Hacking87}

\begin{eqnarray}
\rho_0(L_{60}) &=& 2.94 \times 10^{28} 10^Q,\\
Q &=& Y - \left[ B^2 + 
	\left( {\log L_{60} - X \over  W} \right)^2  \right]^{1/2}
        -2.5 \log L_{60},
\end{eqnarray}
where $B=1.51, W=0.85, X=23.96,$ and $Y=5.93$,
assuming $H_0=100$ km s$^{-1}$ Mpc$^{-1}$ and $q_0=0.5$
(note: in \cite{Hacking87} $Y$ appears to be incorrectly given as 6.93).

The 60 micron luminosity function at redshift $z$ is

\begin{equation}
\rho(L_{60},z) = {g(z) \over f(z)} \rho_0 \left( {L_{60} \over f(z)} \right)
\end{equation}
where $g(z)$ and  $f(z)$ are density and luminosity evolution functions.
According to Gregorich et al. \cite{Gregorich}, $g(z)$ and  $f(z)$
given in table 1 of Condon \cite{Condon84} are consistent with IRAS
source counts.

The intensity (W Hz$^{-1}$ m$^{-2}$ sr$^{-1}$)
from all sources is obtained by integration over
redshift and 60 micron luminosity:

\begin{equation}
I_\nu = {1 \over 4 \pi}
\int dz {d V_c \over dz} {(1+z) \over 4 \pi d_L^2}
\int dL_{60} \, \rho(L_{60},z)
\left( {L_{\nu^\prime} \over L_{1.4}} \right)
\left( {L_{1.4} \over L_{60}} \right) L_{60}
\end{equation}
where $dV_c$ (m$^3$) is the element of co-moving volume.

\subsection{Radio galaxies}

The intensity due to steep-spectrum radio galaxies is obtained in a
similar way using the radio luminosity function at 1.4 GHz

\begin{equation}
I_\nu = {1 \over 4 \pi}
\int dz {d V_c \over dz} {(1+z) \over 4 \pi d_L^2}
\int dL_{1.4} \, \rho(L_{1.4},z)
\left( {L_{\nu^\prime} \over L_{1.4}} \right) L_{1.4}.
\label{eq:I_nu}
\end{equation}

We obtain the radio luminosity function from the visibility function of
``monsters'' obtained by Condon \cite{Condon84} from the Auriemma et
al. \cite{Auriemma} data for elliptical galaxies (visibility function
parameters $B=2.3$, $W=0.75$, $X=26.1$, and $Y=5.47$ for $H_0=50$ km
s$^{-1}$ Mpc$^{-1}$).  Condon \cite{Condon89} shows that the 1.4 GHz
source counts favour strong luminosity evolution by a factor $\sim 16$ at
$z \sim 0.8$.

\begin{figure}
\vspace{16cm}
\includegraphics{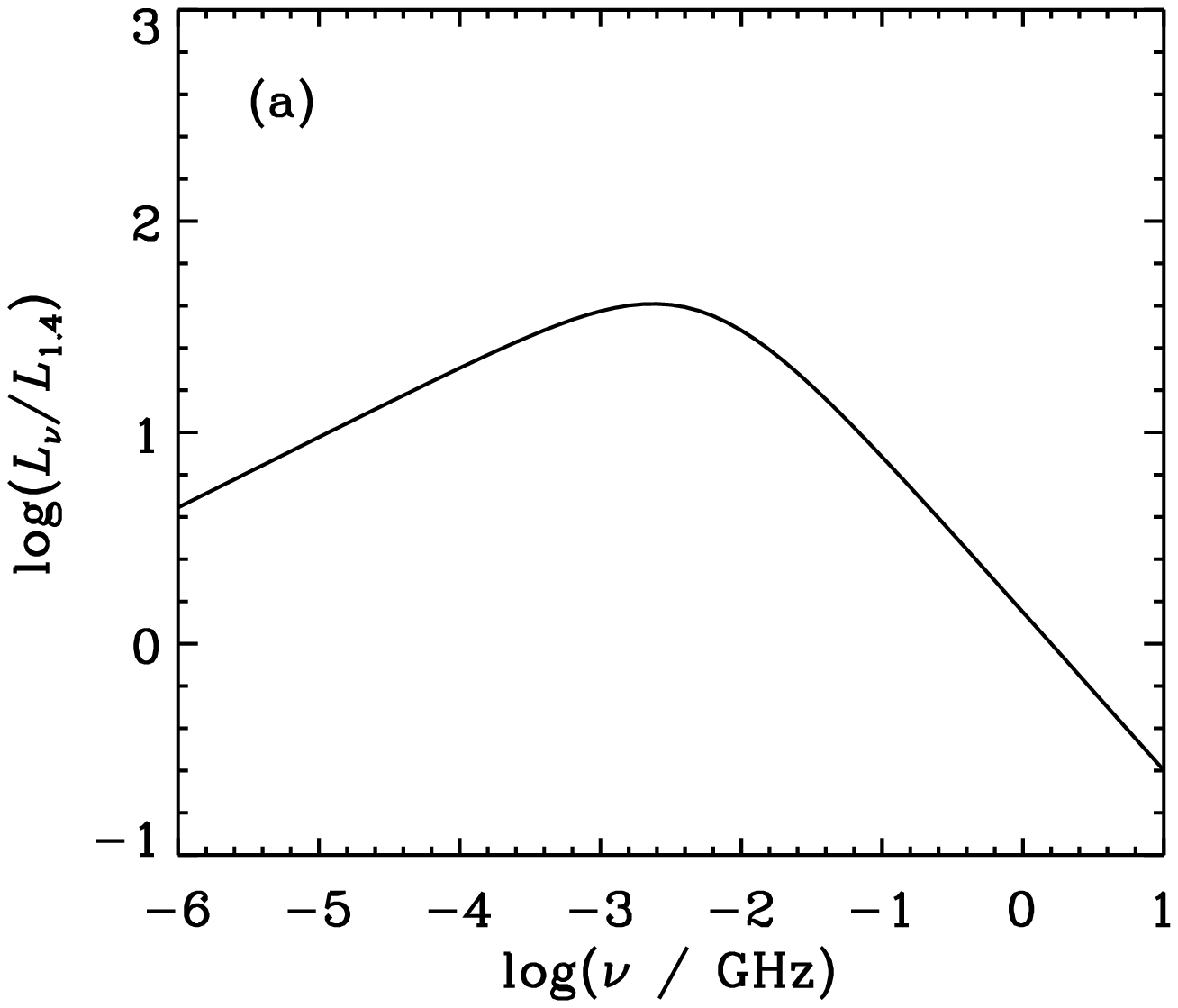}
\includegraphics{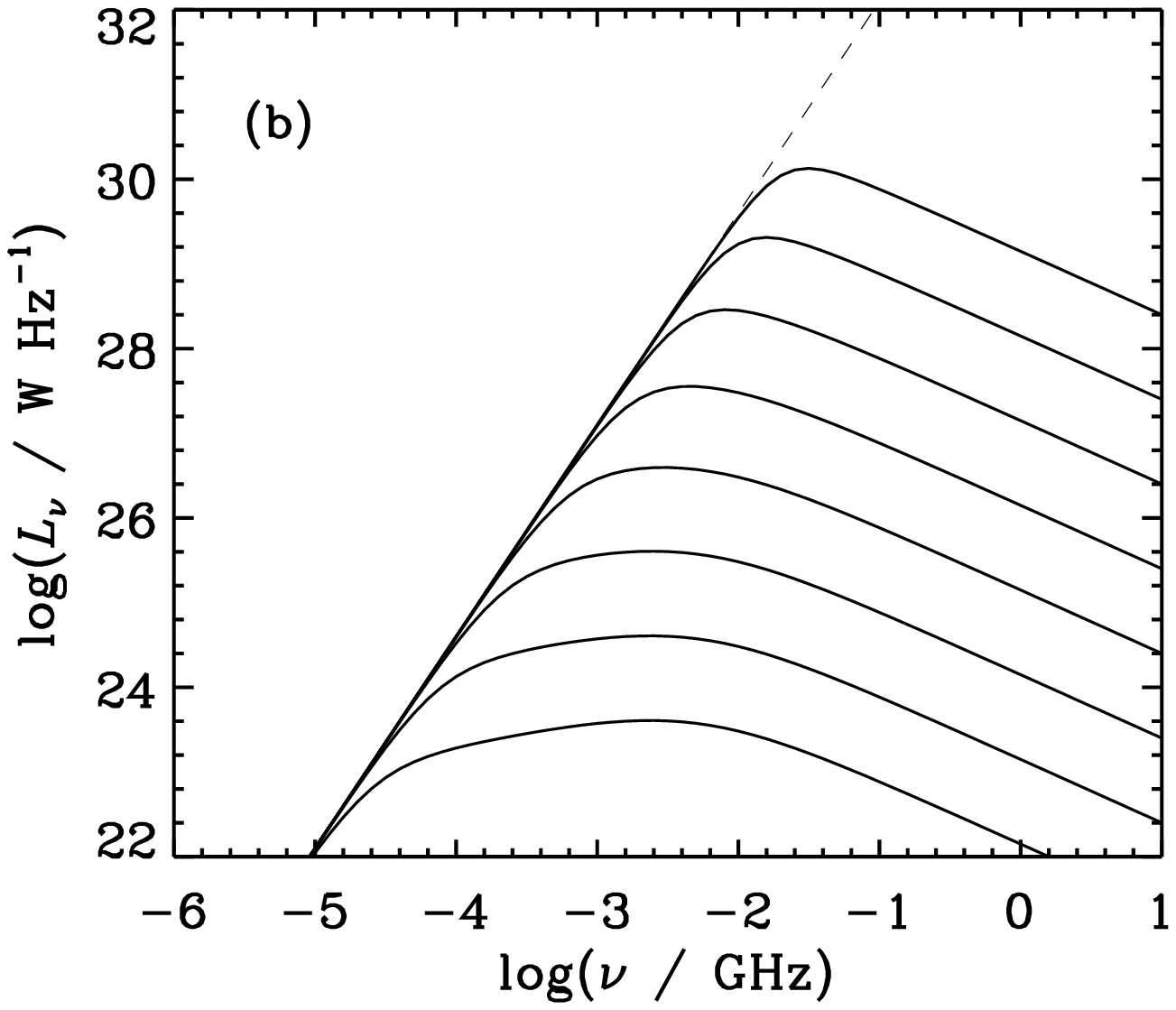}
\caption{(a) Model radio spectrum for radio galaxies neglecting
synchrotron self-absorption.
(b) Average spectra for radio galaxies with luminosity
$L_{1.4} =10^{22}$ \dots $10^{29}$ W Hz$^{-1}$ including the
effects of synchrotron self-absorption.
Dashed line shows spectrum of completely self-absorbed source.}
\label{fig:freefree_rg}
\end{figure}

For the ratio $L_{\nu^\prime}/ L_{1.4}$ we follow Falcke \& Biermann
\cite{Falke&Biermann95} in assuming the electrons responsible for the
synchrotron radiation are $e^\pm$ from $\pi \to \mu \to e$ decay, the
pions being produced as secondaries in interactions of energetic protons
with ambient matter.  Energetic protons may be accelerated in radio
galaxies to a power-law spectrum by shock acceleration
\cite{Biermann&Strittmatter,RB93}.  
We make the usual assumption that the electrons will be in
energy density equipartition with the magnetic field, protons, etc., and
radiate in an environment of typically 10 to 100 microgauss (see
e.g. ref. \cite{Miley80}).  The electron spectrum on production has a
low-energy cut-off at $\gamma_c m c^2 \approx 100$ MeV (see
e.g. \cite{Protheroe82}) giving rise to a break in the synchrotron
spectrum at the break frequency

\begin{equation}
\nu_b = {3 e B_\perp \over 4 \pi m c} \gamma_c^2
\end{equation}
where $\gamma_c \approx 200$ is determined by the pion rest mass and so 
corresponds to the low-energy cut-off of the electron energy distribution.
Adopting a $B_\perp = 30$ microgauss gives rise to a break at $\sim 5$ MHz. 
Below this break frequency the spectrum will be $\nu^{1/3}$ (i.e. the 
asymptotic form of the synchrotron emission coefficient at low frequency for
monoenergetic electrons \cite{Pacholczyk}).  For the spectrum above the break
frequency we use the mean spectral index $\langle \alpha \rangle =0.75$
suggested by Condon \cite{Condon84}. This spectrum is shown in
Figure~\ref{fig:freefree_rg}(a).

Because of the much higher luminosities of radio galaxies compared to
normal galaxies, synchrotron-self absorption will be greater at
low frequencies than in normal galaxies.
Thus the contribution of radio galaxies to the radio background
will be less at low frequencies than that of normal galaxies.
We may therefore use the approximate formula (Equation~\ref{ssa_approx})
with $A_{\rm proj} \sim $ (30 kpc)$^2$ and  $B \sim 10$ microgauss
to obtain the maximum luminosity, and modify the spectrum given in
Figure~\ref{fig:freefree_rg}(a).
The resulting spectra are given in Figure~\ref{fig:freefree_rg}(b) for various
luminosities.

\subsection{Infrared and radio source counts}

There are uncertainties in the evolution of the luminosity functions
of normal galaxies and radio galaxies, in the radio--infrared correlation
used for normal galaxies, and in the models for the radio spectra we used.
We therefore decided to calculate radio and infrared source counts and
compare with observation to ensure that the evolution of the luminosity
functions used in calculating the radio background is consistent with
the source count data.

The number of sources with observed flux
at frequency $\nu$ greater than $S_\nu$ is

\begin{equation}
N(>S_\nu) = \int dz {d V_c \over dz}
\int_{L_{\nu^\prime}}^\infty dL_{\nu^\prime} \, \rho(L_{\nu^\prime},z)
\label{eq:N}
\end{equation}
where

\begin{equation}
L_{\nu^\prime} = {4 \pi d_L^2 \over (1+z)} S_\nu.
\end{equation}
Differentiating equation~\ref{eq:N} with respect to $S_\nu$ we obtain

\begin{equation}
n(S_\nu) \equiv {1 \over 4 \pi}{d N \over d S_\nu} =
\int dz {d V_c \over dz} {d_L^2 \over (1+z)}
\rho(L_{\nu^\prime},z).
\end{equation}

\begin{figure}
\vspace{16cm}
\includegraphics{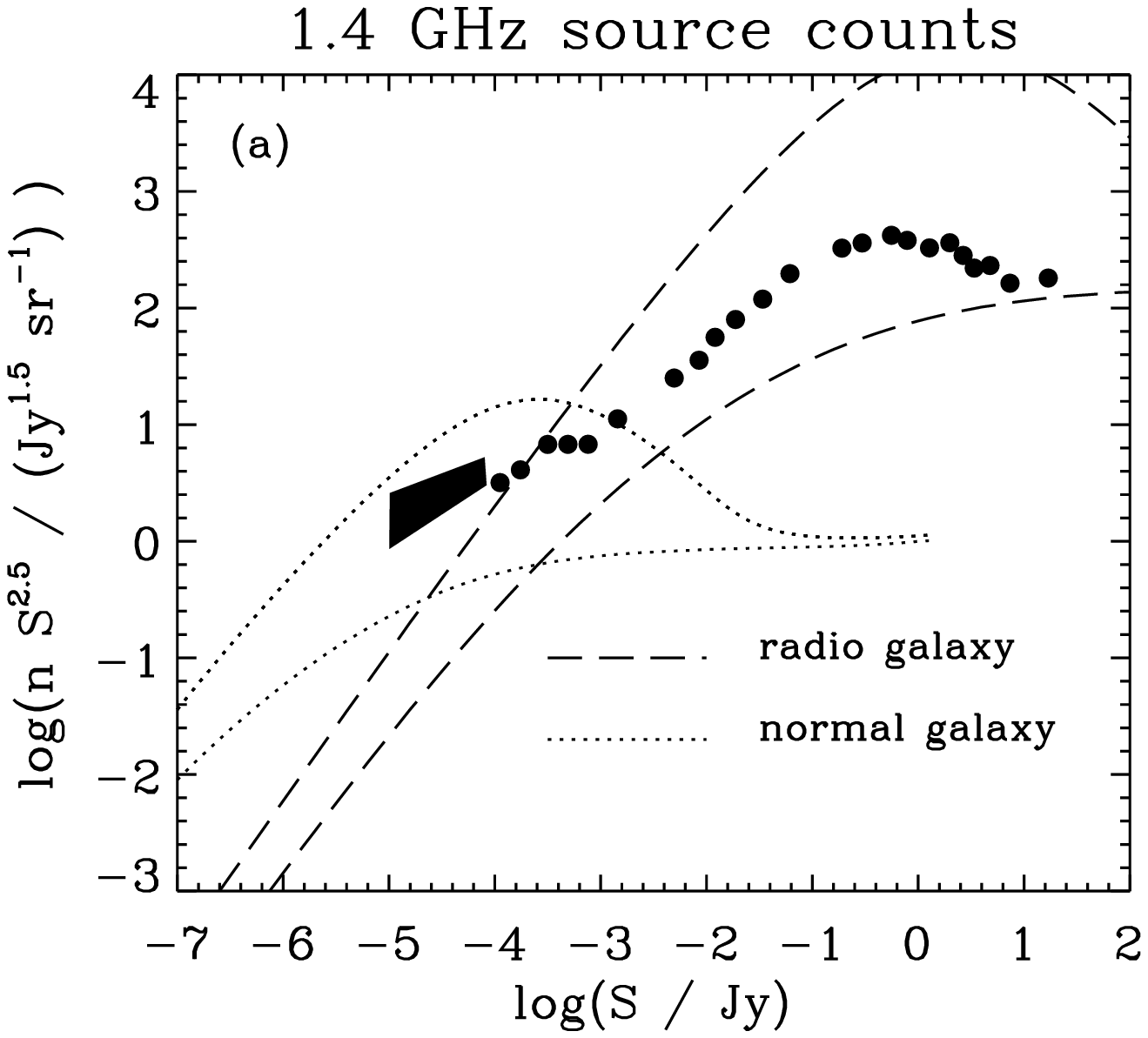}
\includegraphics{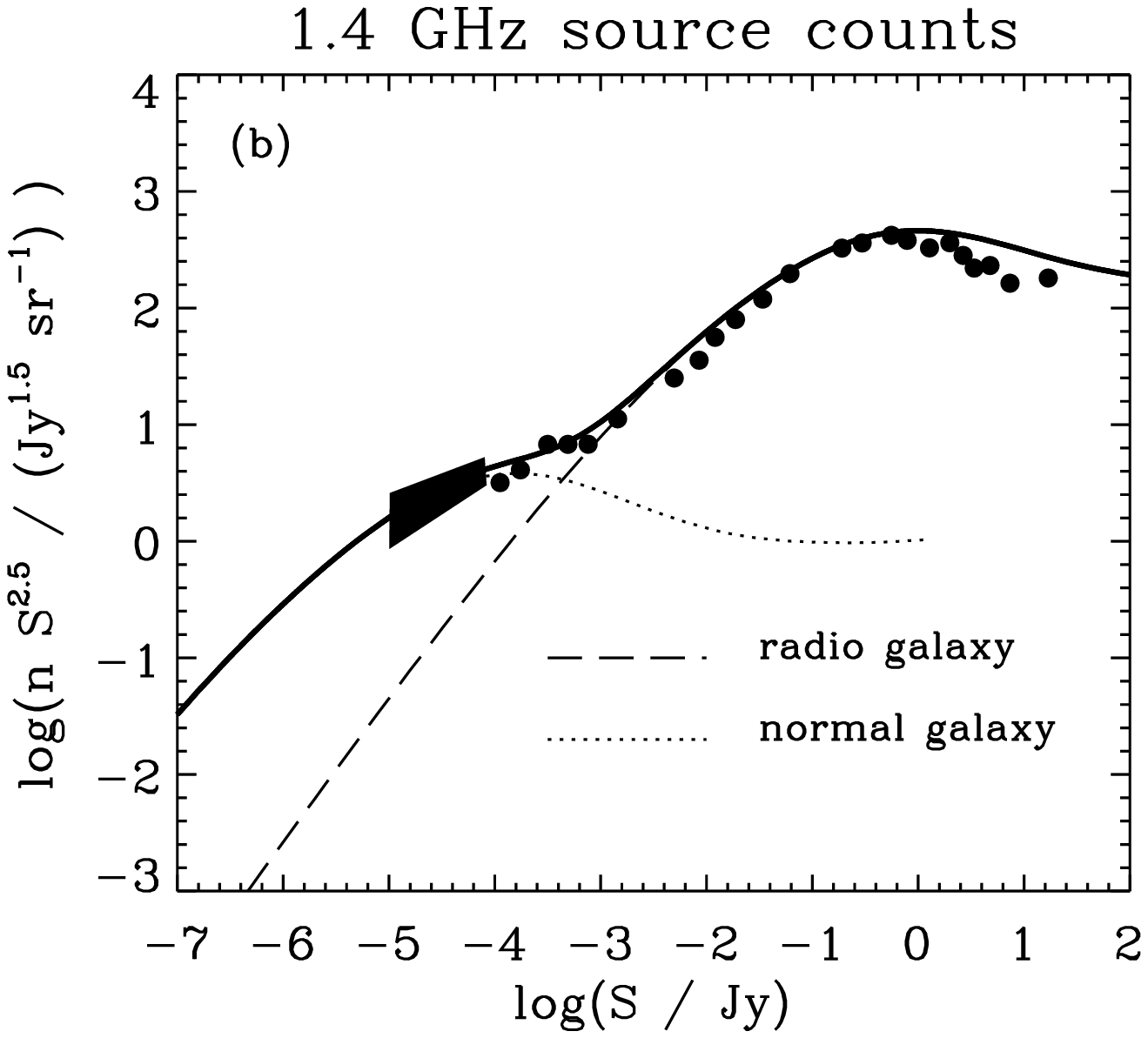}
\caption{Contributions of normal galaxies (dotted curves) and
radio galaxies (dashed curves) to the total extragalactic radio source
counts at 1.4 GHz (solid curve).  (a) Lower curves are for no evolution
and upper curves are for evolution of normal galaxies according to table 1
of Condon \protect \cite{Condon84} and pure luminosity evolution of radio
galaxies described by $(1+z)^4$ at all $z$.  (b) Best fitting pure
luminosity evolution as described in the text; solid curve gives the total
source count.  Data are from ref. \protect \cite{Condon89}.}
\label{fig:radio_sourcecounts}
\end{figure}

We use the same luminosity functions and other input we used in
calculating the radio background to obtain the radio source counts at
1.4 GHz for normal galaxies and radio galaxies.  Resulting radio source
counts of normal galaxies and radio galaxies are shown in
Fig.~\ref{fig:radio_sourcecounts}(a) for the case of i) no evolution and
ii) evolution of the normal galaxy population according to table 1 of Condon
\cite{Condon84}, and pure luminosity evolution of the radio galaxy
population according to $f(z)=(1+z)^4$ at all $z$.  We compare the total
source counts with data obtained by Condon \cite{Condon89}.

\begin{figure}
\vspace{16cm}
\includegraphics{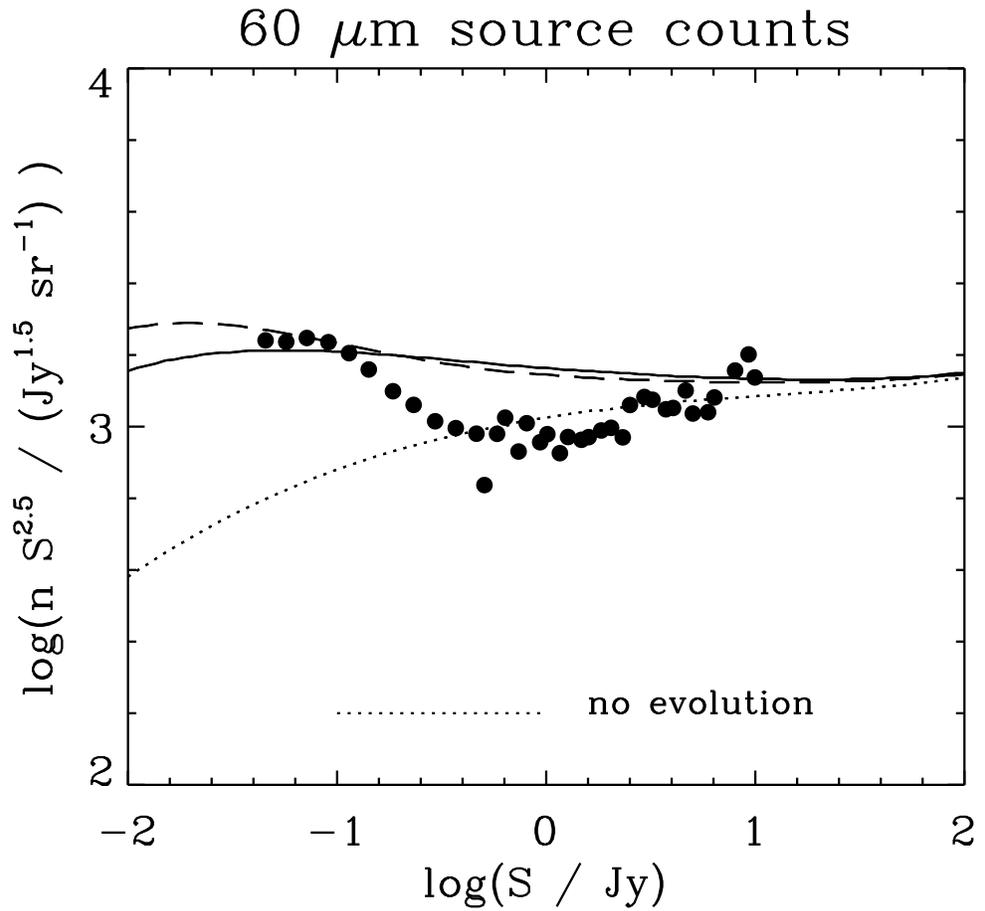}
\caption{Contributions of normal galaxies
to the 60 micron source counts for no evolution (dotted curve), evolution
according to table 1 of Condon \protect \cite{Condon84} (dashed curve),
and with pure luminosity evolution as described in the text (solid curve).
Data are points with error bars smaller than 20\% from the summary in
Figure 2 of ref. \protect \cite{Gregorich}.}
\label{fig:ir_sourcecounts}
\end{figure}

The calculated source counts are dominated by the radio galaxy
contribution at high $S_\nu$ and by normal galaxies at small $S_\nu$.  As
can be seen, in both cases, the predicted source counts fall well below
the data for the case of no evolution and are well above the data for the
evolution models assumed.  We note that the evolution of normal galaxies
according to table 1 of Condon is also approximately pure luminosity
evolution with $f(z)=(1+z)^4$.  Our approach to this problem is to set
$g(z)=1$ and modify the luminosity evolution to fit the source counts.  We
assume luminosity evolution of the form

\begin{equation}
f(z)= \left\{ \begin{array}{ll} (1+z)^4 & z < z_0 \\
                                (1+z_0)^4 & z \ge z_0,
        \end{array} \right.
\end{equation}
and we adjust $z_0$ to fit the data.  We
find the best fit is given by $z_0=0.8$ for both normal galaxies and radio
galaxies.  The resulting radio source counts of normal galaxies and radio
galaxies are shown in Fig.~\ref{fig:radio_sourcecounts}(b) and are seen to
be in good agreement with the data.  As a further check, we have
calculated the 60 micron source counts due to normal galaxies for the case
of no evolution, evolution of the normal galaxy population according to
table 1 of Condon \cite{Condon84}, and the evolution model described above
which fits best the radio source counts.  The results are shown in
Figure~\ref{fig:ir_sourcecounts} where we see that the evolution model we
adopt gives as good a fit as table 1 of Condon \cite{Condon84}.

We note that by omitting data with statistical errors larger than 20\% in
Figure~\ref{fig:ir_sourcecounts} (data with large error bars were included
in the original plot of Gregorich et al. \cite{Gregorich}) we see that for
fluxes between 0.3 and 3 Jy both models predict source counts which are
higher than the data.  Whether this is an indication of a new source
population contributing at $S_\nu < 0.3$ Jy (perhaps AGN),
or some systematic effect affecting the data, remains to
be seen.  However, Gregorich et al. \cite{Gregorich} argue that their data
favour the evolution model of Condon \cite{Condon84} rather than no
evolution, and our adopted evolution model gives as good a fit to these data
as Condon's.  We therefore proceed to calculate the mean free path for
interactions of $\gamma$-rays in our calculated radio background
for two possible cases: (a) no evolution of the normal galaxy population
(we assume a new source population below 0.3 Jy which does not contribute
significantly to the radio background); and (b) evolution of the
normal galaxy population (we assume the sources with $S_\nu < 0.3$ Jy
are still normal galaxies).

\section{The radio background and interactions of $\gamma$-rays}

\begin{figure}
\vspace{16cm}
\includegraphics{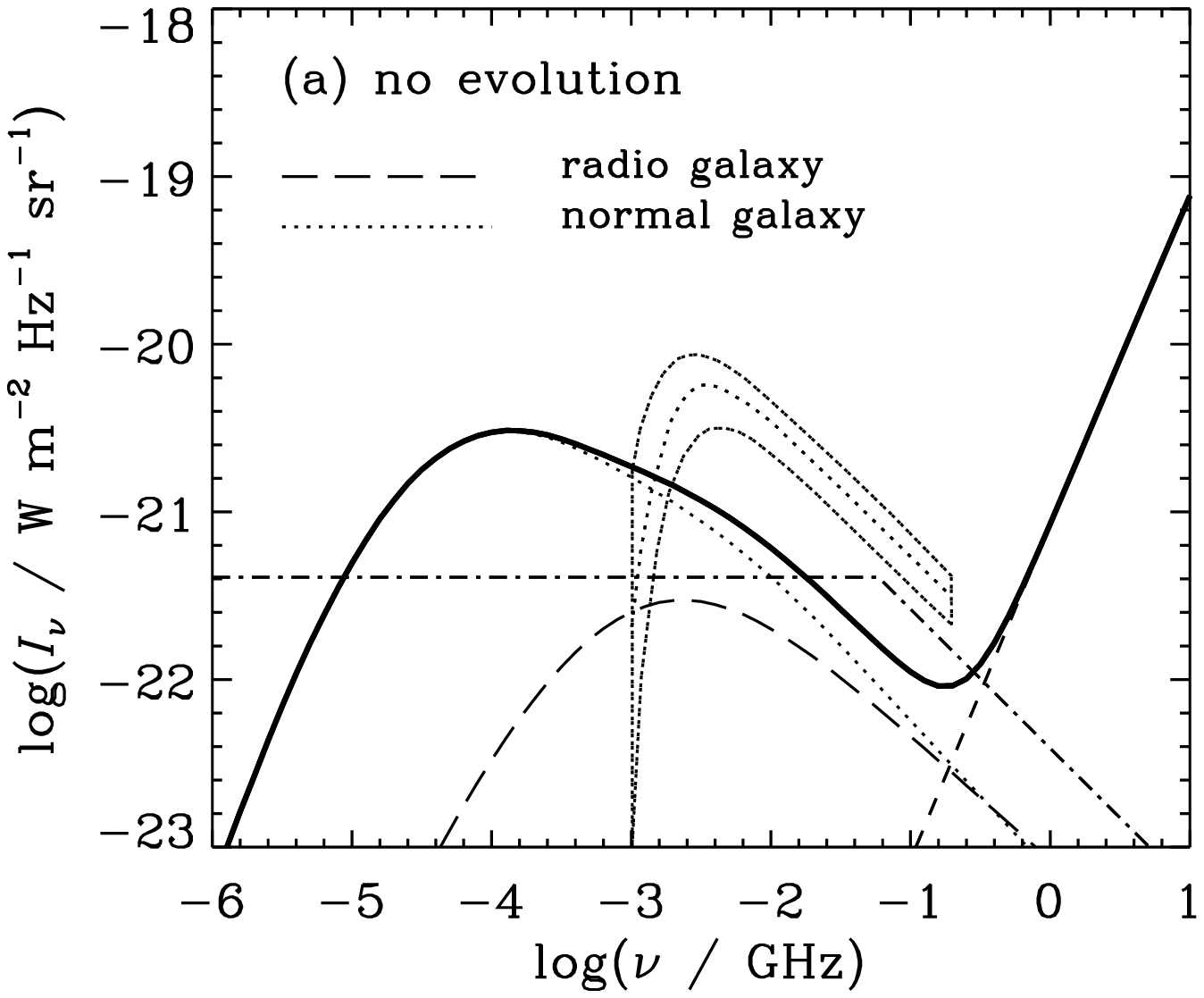}
\includegraphics{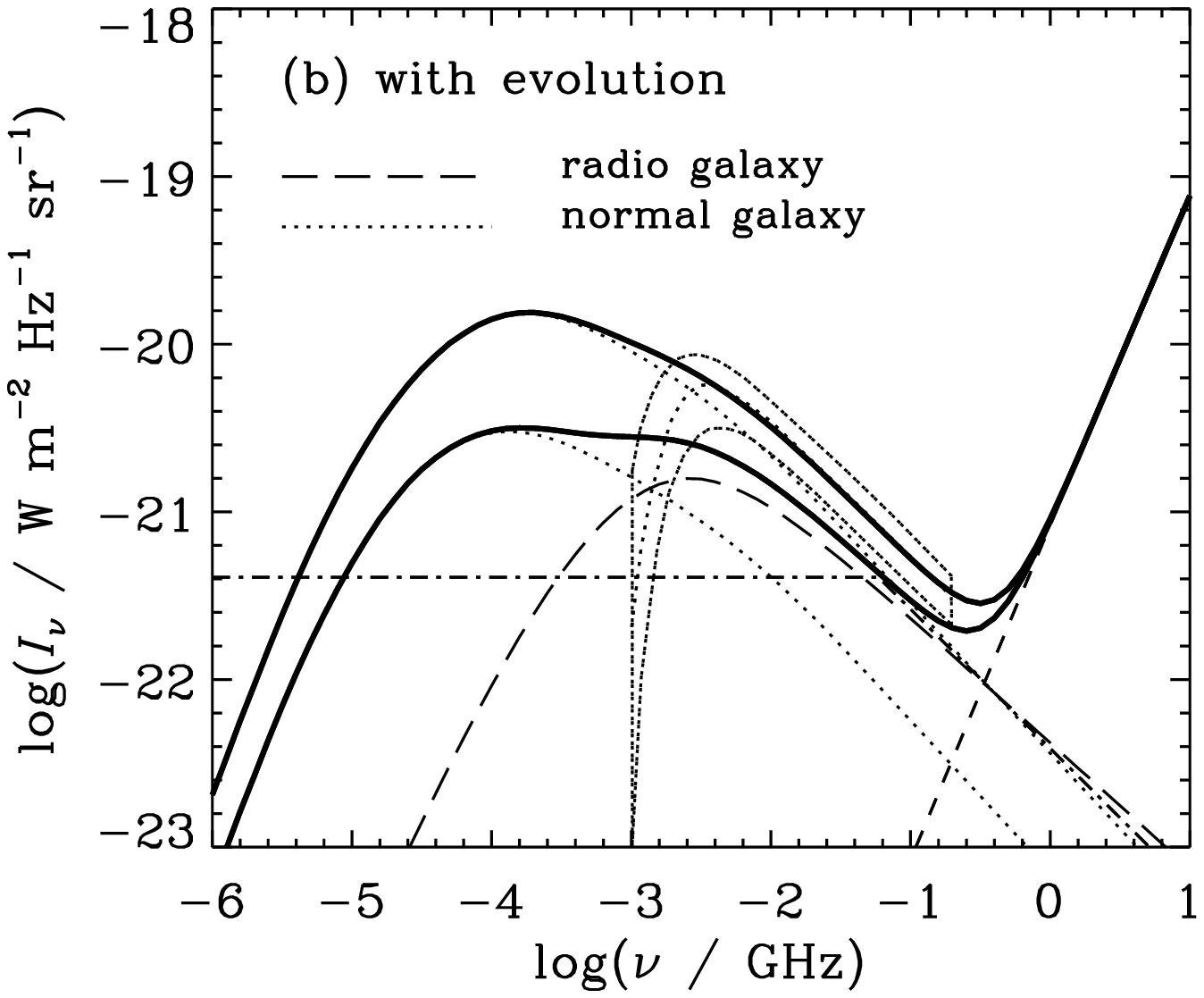}
\caption{Contributions of normal galaxies (dotted curves),
radio galaxies (long dashed curve),
and the cosmic microwave background (short dashed curve)
to the extragalactic radio background intensity (thick solid curves)
for (a) no evolution and (b) with pure luminosity evolution as
described in the text (upper curves), and with pure luminosity evolution
only for radio galaxies (lower curves).
Dotted band give an observational estimate of the total
extragalactic radio background intensity \protect \cite{Cla70}
and the dot-dash curve gives an earlier theoretical estimate
\protect \cite{Ber69}.}
\label{fig:radio_background}
\end{figure}

The contributions to the extragalactic radio background intensity from
normal galaxies, radio galaxies, and the cosmic microwave background,
together with the total estimated radio background intensity are plotted
in Figure~\ref{fig:radio_background}.  Our result is compared with the
total extragalactic radio background intensity estimated from observations
by Clark et al. \cite{Cla70} and with a theoretical estimate made several
years ago by Berezinsky \cite{Ber69}.  We note that with our adopted model
for the radio spectrum from normal galaxies, the normal galaxies 
dominate the background as suggested by their
dominance in source counts at low flux density levels (\cite{BEW}).

Photon-photon interactions are described in \cite{Prot86} and
references therein.
The mean interaction length, $\lambda$, of a photon of
energy $E$ is given by,

\begin{equation}
        [\lambda(E)]^{-1}= {1 \over 8 E^2}
\int_{\varepsilon_{\rm min}}^{\infty} \, d\varepsilon \frac{n(\varepsilon)}
        {\varepsilon^2} \int_{s_{\rm min}}^{s_{\rm max}(\varepsilon,E)}
        ds \, s \sigma(s),
        \label{eq:mpl}
\end{equation}
where $n(\varepsilon)$ is the differential photon number density
of photons of energy $\varepsilon = h \nu$,

\begin{equation}
n(\varepsilon) = {4 \pi \over h c} {I_\nu \over h \nu},
\end{equation}
and
$\sigma(s)$ is the total cross section for photon-photon pair production
\cite{Jauch}
for a centre of momentum frame energy squared given by

\begin{equation}
s=2 \varepsilon E(1 - \cos \theta)
\label{eq:s}
\end{equation}
where $\theta$ is the angle between the directions of the
energetic photon and the background photon, and

\begin{eqnarray}
s_{\rm min} &=& (2 m_e c^2)^2,\\
\varepsilon_{\rm min} &=& {(2 m_e c^2)^2 \over 4E},\\
s_{\rm max}(\varepsilon,E) &=& 4\varepsilon E.
\end{eqnarray}

\begin{figure}
\vspace{12cm}
\includegraphics{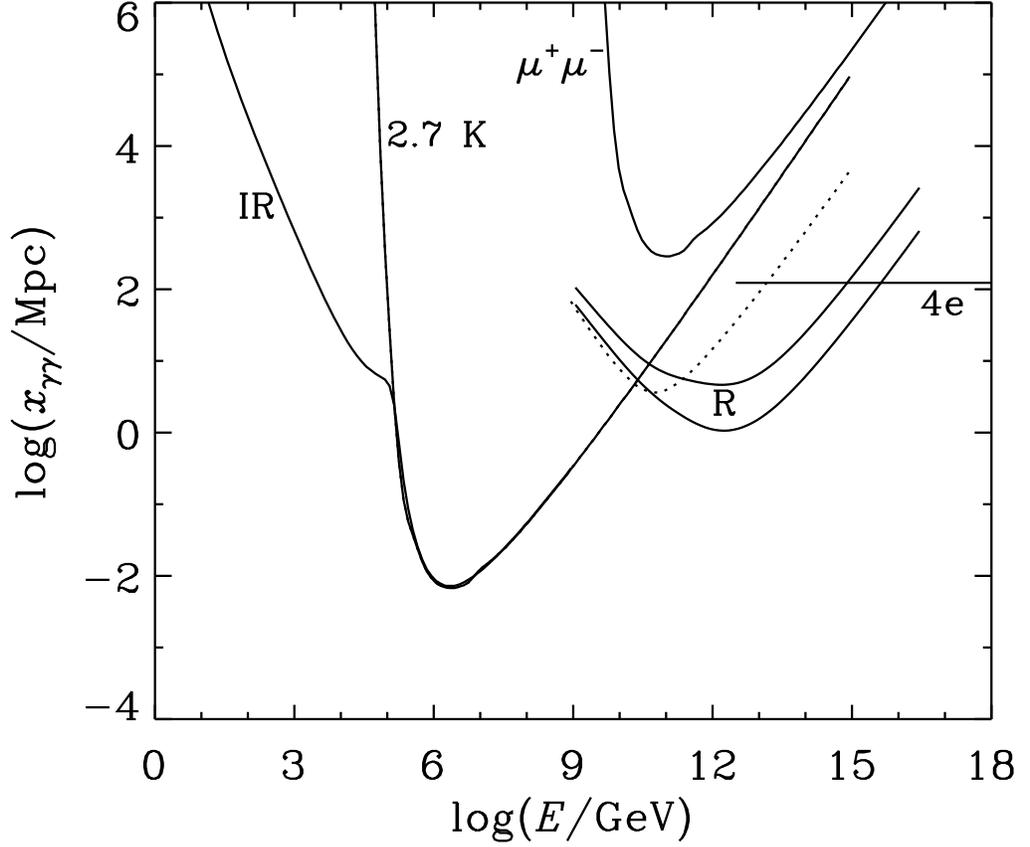}
\caption{The mean interaction length for pair production for
$\gamma$-rays in the Radio Background calculated in the present work
(solid curves labelled R:  upper curve -- no evolution of normal galaxies;
lower curve -- pure luminosity evolution of normal galaxies)
and in the radio background of Clark \protect
\cite{Cla70} (dotted line).  Also shown are the mean interaction length
for pair production in the microwave background (2.7K), the infrared and
optical background (IR), and muon pair production ($\mu^+\mu^-$) and
double pair production (4e) in the microwave background
\protect \cite{ProtJohns95}.}
\label{fig:ggee_radio.eps}
\end{figure}

The interaction length for photon-photon pair production in the
radio background is plotted in Fig.~\ref{fig:ggee_radio.eps}
along with those for competing processes and other radiation fields
\cite{ProtJohns95}.
We also show the mean interaction length for the radio spectrum based on
direct observations together with attempts at subtraction of the effects
of galactic absorption and background \cite{Cla70}.

\section{Conclusion}

Motivated by a new interest in electromagnetic cascades
through the universe at extremely high energies,
we have made a new calculation of the extragalactic radio background
radiation down to kHz frequencies.
The main contribution to the background is from normal galaxies and is
uncertain due to uncertainties in their evolution.
The 60 micron source counts from IRAS above 0.3 Jy appear consistent with no
evolution provided there is a new source population
(possibly AGN) contributing below 0.3 Jy.
An alternative interpretation of the data is that there is strong evolution
of normal galaxies giving agreement with the source counts above 3 Jy and
below 0.1 Jy (but not for $0.1 < S_\nu < 3$ Jy).
This gives rise to a factor of 5 uncertainty in the radio intensity at
kHz frequencies, and this translates to a factor of 5 uncertainty in the
mean free path at $10^{12}$ GeV.
If there is a new source population contributing to the infrared source
counts it may also be important in determining the infrared background
which limits the transparency of the universe to TeV energy gamma rays.
Clearly, it is vital to determine the nature of the sources which
dominate the 60 micron counts below 0.3 Jy.

We calculated the radio background for the two assumptions about the
evolution of normal galaxies, and in both cases the background we
obtain exceeds previous estimates at low frequencies.
By examining Fig.~\ref{fig:ggee_radio.eps}
we find that for the radio background calculated in this paper
photon-photon pair production on the radio background
is the dominant interaction process
for photons over four or five decades of energy from
$3 \times 10^{10}$ -- $5 \times 10^{10}$ GeV to 
$10^{15}$ -- $5 \times 10^{15}$ GeV, above which
double pair production on the microwave background dominates.
We estimate the mean free path to be $\sim 1$ -- 5 Mpc at $10^{12}$ GeV.
Using the radio background estimated by Clark \protect \cite{Cla70}
photon-photon pair production on the radio background
would only be important only up to $10^{13}$ GeV, and the
mean free path at $10^{12}$ GeV would be a factor of 3 -- 10 larger.
This difference will be very important in electromagnetic cascades initiated
by particles with energies up to the GUT scale produced at topological
defects.

\section{Acknowledgment}

P.L. Biermann wishes to thank Reinhard Schlickeiser and
Andy Strong for intense discussions of the cosmic ray electrons in the Galaxy.
We also thank Bram Achterberg, Tom Gaisser, Hinrich Meyer and
Matthew Whiting for detailed comments on the manuscript, and 
Henning Seemann for help with the preparation of the paper. 
The research 
of R.J. Protheroe is funded by a grant from the Australian Reseach Council.

\end{document}